\documentclass[conference, letterpaper]{IEEEtran}
\usepackage{graphicx} % Required for inserting images
\usepackage[utf8]{inputenc}
\usepackage{apacite}
\usepackage[margin=1in]{geometry}
\usepackage{fancyhdr}
\usepackage{cancel}	
\usepackage[english]{babel}
\usepackage[cmex10]{amsmath}
\usepackage{amsmath,amssymb,amsfonts}
\usepackage{algorithmic}
\usepackage{textcomp}
\usepackage{xcolor}
\usepackage{fontawesome}
\usepackage{url}
\usepackage{adjustbox}
\usepackage{array}
\usepackage{booktabs}
\usepackage{listings}
\usepackage[caption=false,font=footnotesize]{subfig}

\renewcommand{\thispagestyle}[2]{} 

\fancypagestyle{plain}{
        \fancyhead{}
        \fancyhead[C]{first page center header}
        \fancyfoot{}
        \fancyfoot[C]{first page center footer}
}
\begin{document}

\title{\textbf{CATASTROAGRI} - Interactive data analysis and visualization application with a future projection for catastrophic agricultural insurance.}

\author{
\IEEEauthorblockA{Marizol Lizbeth Serrano Quispe}
\IEEEauthorblockA{Ingenieria Estadistica e Informatica\\
Universidad Nacional del Altiplano de Puno,Peru.\\
Email: mserranoq@est.unap.edu.pe}
\and
\IEEEauthorblockA{Torres-Cruz Fred}
\IEEEauthorblockA{Ingenieria Estadistica e Informatica\\
Universidad Nacional del Altiplano de Puno ,Peru.\\
Email: ftorres@unap.edu.pe}
}

\maketitle

\IEEEpeerreviewmaketitle
\begin{abstract}
CATASTROAGRI is an application developed to load, analyze and interactively visualize relevant data on catastrophic agricultural insurance. It also focuses on the analysis of an ARIMA (0,1,1) (0,1,1) model to identify and estimate patterns in the agricultural data of the Puno Region, it presents a decreasing trend because there is a significant relationship between successive values of the time series, We can also state that it is not stationary because the mean and variance do not remain constant over time and the series has periods, and it is observed that the cases are decreasing and increasing over the years, especially the amount to indemnify due to the behavior of the climate in the highlands. 
The results of the analysis show that agricultural insurance plays an important role in protecting farmers against losses caused by adverse climatic events. The importance of concentrating resources and indemnities on the most affected crops and in the provinces with the highest agricultural production is emphasized.
The results of the users' evaluation showed a high level of satisfaction, as well as ease of use.
\end{abstract}

\textbf{Keywords -} CATASTROAGRI, agricultural insurance, data analysis, visualization, time series, ARIMA, forecasting, Puno.

\section{INTRODUCTION }
Agricultural activities are important for the overall development of the ancient culture of any nation for thousands of years\cite{bula2020importance}. In terms of healthy food, agricultural activities influence the energy needs of human beings. Any crop goes through three fundamental phases: planting, monitoring, and harvesting. Each of these phases involves a series of activities\cite{wakchaure2023application}.
Today, agriculture is crucial to the survival of the world's population. It can also be said that nations capable of cultivating plants and raising animals in different ecological conditions have control over food power. It is well known that most agricultural research is carried out in developed countries, but it is important to determine the extent to which it is carried out in Latin America and the Caribbean.\cite{chura2019necesidades}
High-performance technologies allow us to generate large amounts of data that could be used for research in any of the following fields\cite{anvzel2022movis}.In the Puno Region, the Direccion De Estadistica Agraria e Informatica; faces a series of challenges year after year, one of them being catastrophic risks that could negatively affect crop production and the livelihood of farmers. catastrophic agricultural insurance has been implemented since 2009 consecutively in each agricultural campaign.\cite{vinelli2015informe} 
The Regional Agricultural Directorate of Puno, a dependency of the Regional Government of Puno, has developed actions of Supervision and Monitoring of Catastrophic Agricultural Insurance, during the agricultural campaigns year after year, cooperating with the Technical Secretariat of FOGASA of MINAGRI in the insurance process and in the evaluation of the performance of the Insurance Company RIMAC SEGUROS. It coordinates the implementation, evaluations and adjustments in the field, tasks that are carried out from loss notices generated by the agricultural agencies and entered into the system provided by RIMAC Seguros, virtually. It assumes the risk, pays compensation to farmers in the event of a catastrophic weather event that causes the total or near-total loss of the crop and aims to reduce the vulnerability to which the crops of small farmers with limited resources are exposed, especially those of subsistence family farming\cite{olascoaga2023gestion}. \\
However, to maximize the benefits of the SAC and encourage more efficient management, a tool is required to analyze and visualize insurance-related data in a clear and accurate manner.  An interactive application has been created using the Python programming language in conjunction with Streamlit, an open-source framework for creating interactive data-driven web-based applications\cite{chaure2022forecaster}, to respond to this demand and offer a complete platform for studying and making decisions on catastrophic agricultural insurance.\\
This application not only allows to load and process relevant data, but also allows access to basic statistics, graphs of the information required, and filtering of detailed data at different geographical levels, such as provinces, districts, statistical sectors, particular crops, and agricultural campaigns. 
\section{METHODS AND MATERIALS} 
\subsection{Study design}
The study was carried out through observation and analysis.  Data was collected from the Regional Agrarian Directorate of Puno on the list of producers insured with the agrarian insurance and statistical analyses were carried out to evaluate the indemnity of insured areas\cite{rodriguez2017factores}.
\subsection{Study population}
The study population is the 485 statistical sectors where farmers in the Puno region are registered in the official register of catastrophic agrarian insurance.

\begin{figure}[h!]
    \centering
    \includegraphics[width=8cm,height=8cm]{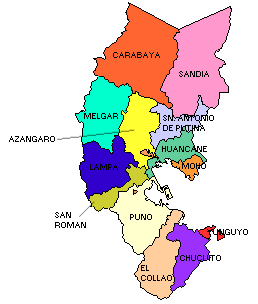}
    \caption{Direccion Estadistica Agraria Informatica}
\end{figure}

\subsection{Data collection}
The data used in this study were acquired from the Direccion Regional Agraria Puno, from the office of Direccion Estadistica Agraria Informatica and from the Fondo de Garantia para el Campo y del Seguro Agrario (FOGASA)\cite{falconi2017situacion}.\\ This data includes information on:\\
\begin{table}[htbp]
    \begin{flushleft}
    \begin{adjustbox}{max width=3\textwidth}
    \begin{tabular}{|c|c|c|}
        \hline
         Variable & Description & Type of Variable \\
        \hline
        Year & Campaña Agricola & Char \\
         Province & 14 Provinces & Char \\
         District & 110 Districts & Char \\
         Statistical Sector & 485 Sectors E. & Char \\
         Crop Name & Papa, Quinoa, etc. & Char \\
         Sown Area (Has) & & Int \\
         Insured Area (Has) & & Int \\
        Amount Ind (Soles) & & Int \\
         N. of prod benefited & & Int \\
        \hline
    \end{tabular}
    \end{adjustbox}
    \end{flushleft}
    \caption{Description of variables}
\end{table}

\subsection{Data analysis}
A statistical analysis of the data collected was carried out using the Python programming language and various libraries, such as Pandas and NumPy. We calculated the total insured area per province, total sown area per province, district, statistical sector, indemnity amount per crop, province, district, etc. With the results, we can visualize graphs showing the totals in types of graphs such as: bar, line, histogram, scatter, and circular, and finally a prediction for the next agricultural campaign with respect to the insured area and the amount to be indemnified.

\subsection{Application development}
The interactive application was developed using the Streamlit framework in conjunction with Python\cite{coppiano2022desarrollo}. Different functions and algorithms were implemented to load, process, and visualize data related to agricultural catastrophic insurance. Interactive graphs, tables, and filters are included to facilitate user data exploration and analysis.

{Pseudocodigo}

\lstdefinestyle{pseudocode}{
  backgroundcolor=\color{gray!10},
  basicstyle=\footnotesize\ttfamily,
  numbers=left,
  numberstyle=\tiny\color{gray},
  stepnumber=1,
  numbersep=5pt,
  keywordstyle=\color{blue}\bfseries,
  commentstyle=\color{green!60!black},
  stringstyle=\color{red},
  showstringspaces=false,
  tabsize=2,
  breaklines=true,
  breakatwhitespace=true,
  frame=single,
  captionpos=b,
  aboveskip=9pt,
  belowskip=9pt,
}

\begin{lstlisting}[style=pseudocode, caption={Pseudocódigo}]
Import the necessary libraries

def load_data(file):
    if file extension is CSV:
        load data from CSV file
    or if the file extension is Excel:
        load data from Excel file
    else:
        display a warning message indicating that the file format is not supported
        return None
    return the loaded data
def parse_data(data):
    perform the analysis of the data according to the requirements
    display the results of the analysis

def visualize_data(data):
    perform the visualization of the data according to the requirements
    display the generated charts or visualizations
def download_charts_in_excel(data):
    define the name and location of the output file
    save the data to an Excel file in the specified location
    display a success message indicating the path to the downloaded file
def main():
    set the application page
    load a header image
    display the application title and header
    display options to load a file and select functionality
    if a file is loaded:
        load data from file
        if "Data analysis" functionality is selected:
            if the data was loaded successfully:
                analyze the data
        if "Display data" functionality is selected:
            if the data was loaded successfully:
                display the data
        display a button to download the tables to Excel
    if the download button is clicked:
        if the data was loaded successfully:
            download the tables in Excel
            TEMPORARY SERIES
    run the application by calling the main function

\end{lstlisting}

\begin{figure}[h!]
    \centering
    \includegraphics[width=8cm,height=12.5cm]{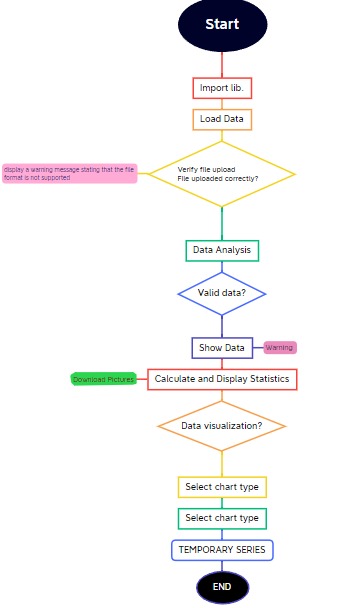}
    \caption{Flowchart }
\end{figure}

The Streamlit library is used to create an interactive web interface that allows data loading, analysis, and visualization\cite{dominguez2022analisis}. Streamlit is a popular library for rapid web application development in Python\cite{Streamlit}, particularly in the field of data science. 
The code is based on functions such as load data for loading data from a file, analysis data for analyzing data and creating charts for creating interactive graphs. Other libraries, such as Pandas for data management, Plotly Express for visualization, and PIL for image processing, offer these functions\cite{pineda2023diseno}. Overall, the code provides an easy-to-use web interface for exploring and understanding the data\cite{McKinney:2010}.

\begin{itemize}
    \item Imports: The code imports libraries such as Streamlit for the interface, Pandas for data management, Plotly for visualizations, and others necessary for the application.
\item Page configuration: sets the title, icon, and page layout.

 \item The "data loader" function allows loading a data set from an Excel file.

 \item The analysis and visualization of loaded data have several defined functions.

 \item User interface: Streamlit was extracted to create the user interface, which includes options for loading data, choosing columns for analysis and visualization, and making time-series predictions.
 \item User interface: Streamlit was extracted to create the user interface, which includes options for loading data, choosing columns for analysis and visualization, and making time series forecasts.

 \item Time series forecasts: The loaded data is used to make time series forecasts using a linear regression model.

 \item Download files: An option is added that allows downloading the data that has been analyzed in an Excel file.
\end{itemize}

In summary, the code creates an interactive application that uses Streamlit and various Python libraries to load, analyze, visualize, and predict data in Excel format. The application can make future time-series predictions and analyze and gain insights from the loaded data.

\subsubsection{Application validation}

The application will be validated by exhaustive testing to ensure its correct functioning and the accuracy of the results presented\cite{carlos2022aplicativo}.
\section{RESULTS  }

\begin{figure}[h!]
    \centering
    \includegraphics[width=8cm,height=6cm]{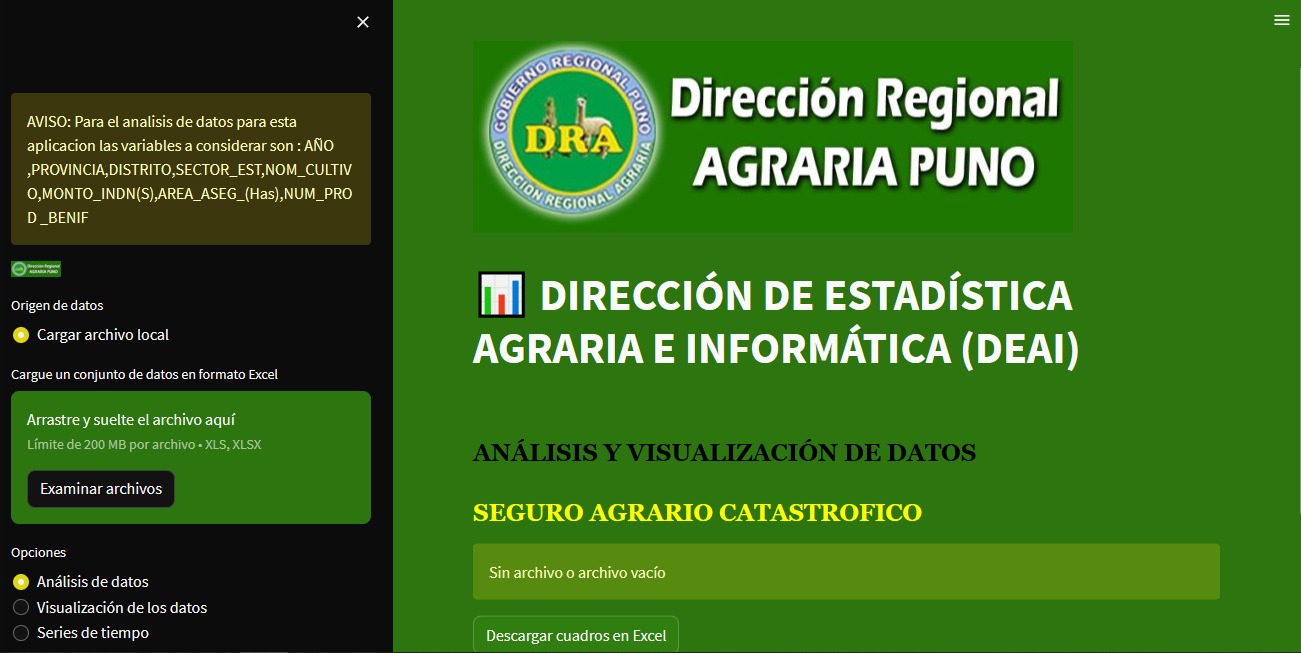}
    \caption{AGROSEGURO interface }
\end{figure}
In this part, the results of how the application interacts with the user, a specialist in the topic of Catastrophe, together with a producer's register, in this case, would be the Office of Agricultural Statistics and Informatics.
\subsection{Data analysis:}
AGROSEGURO is an application that is designed for the statistics area, because they handle a large amount of data, if you want to make a deep analysis the most used and known tool is Excel, but for its manipulation, one spends more than the desired time to get summary tables.
When loading the data, Figure 4 shows the loaded data and the variables needed for the final consolidation.

\begin{figure}[h!]
    \centering
    \includegraphics[width=8cm,height=5cm]{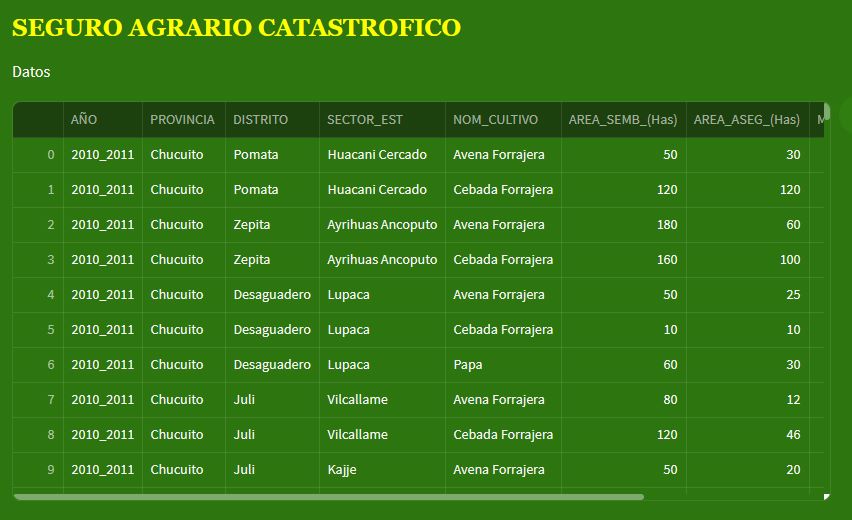}
    \caption{Interfaz de Aplicación SAC en Streamlit}
\end{figure}

The term user-friendly refers to a hardware device or software interface.  These devices or interfaces are said to be "user friendly" because it means that they are not difficult to learn or understand, which leads to user-friendly interfaces\cite{castillo2018diseno}.

Basically what the application does is to visualize in a quick and simple way the results that we need to get the statistics, when loading the desired file of the agricultural campaign to know, this must contain several specific and necessary variables for the final consolidated as: Year of the agricultural campaign (YEAR), Province it is known that Puno currently has 110(PROVINCE),(DISTRICT), Statistical Sector, Puno has 485 (SECTOR EST), Name of Crop, such as potato, quinoa, forage oats, (NOM CROP), Amount to compensate the producer (AMOUNT INDN(S)), Insured Area (AREA ASEG(Has)), Registered planted area (AREA SEM(Has)), finally the number of benefited producers (NUM PROD BENIF).\\
Figure 4 shows the data loaded for later analysis.

\textbf{CLIMATIC RISKS:}\\
The risks covered by catastrophic agricultural insurance are of climatic and biological origin\cite{romero2017seguro}:\\
\textbf{Climate-related risks:}\\
 Drought, excessive or unseasonal rain, landslides, floods, lack of soil for harvesting, excessive humidity, frost and low temperature, hail and snow, high temperatures, and strong wind.\\
 \textbf{Risks of biological nature:}\\
 Pests and predators, diseases.\\
 \textbf{Others;}\\ Fire, volcanic eruption, earthquake.
The procedure to benefit producers:
\begin{enumerate}[]
\item 	Once the climatic event occurs, the producer notifies the agrarian agency or agrarian office, or the agrarian agent becomes aware of the climatic events that have occurred in the area.
\item 	The agricultural agency or office, after evaluation, enters the data of the damages and losses caused by climatological phenomena in the SISAGRI System of RIMAC Seguros, in virtual form, the data of the occurrence is entered by statistical sector, has affected and lost, type of phenomenon, vegetative period, date and is disappointed by the insurance company, with this procedure has been fulfilled in the notice of loss.
\item 	RIMAC SEGUROS sends personnel to perform the respective field evaluation and adjustment.
\item 	If the adjustment report indicates a catastrophic loss, the insurance company RIMAC will pay S/.800 soles for each insured hectare of the insured crop in the statistical sector.
\end{enumerate}
The adjuster appointed by the insurance company carries out the inspection, crop estimation, and loss adjustment upon receipt of the notice of loss, accompanied by personnel from the agricultural agencies, signing the adjustment report and field evaluation.\\
If the adjustment report determines that there is a catastrophic loss of the insured crop in the Statistical Sector, RIMAC SEGUROS indemnifies the agricultural producer, upon receipt of the adjustment report, the amount equivalent to:

\begin{gather}
   COMPENSATION:(S/.) = 800 * SACASE
\end{gather}

Where:
SACASE: Insured area of the insured crop in the Statistical Sector.\\
\begin{table}[ht]
    \centering
    \begin{tabular}{|c|c|c|c|}
        \toprule
        CULTIVO & {SUP-ASEG(hs)} & {Mnt INDM} & {N.PROD BEN} \\
        \midrule
        Avena For & 4948 & 1979200 & 16536 \\
        Cebada Gra & 3295 & 1318000 & 14092 \\
        Haba & 1338 & 535200 & 6531 \\
        Papa & 2650 & 1060000 & 4294 \\
        Quinua & 2607 & 1042800 & 6476 \\
        \midrule
        TOTAL & 14838 & 5935200 & 47929 \\
        \bottomrule
    \end{tabular}
    \caption{REGION PUNO: CATASTROPHIC AGRICULTURAL INSURANCE BY CROP, 2010-2011 CROP YEAR}
\end{table}

The amount of cash-receiving indemnity has a positive confirmation with the insured area, indicating that compensation is distributed according to the losses suffered.
It was found that the crops of fodder oats and grain barley account for most of the compensation in the Puno region when analyzing the amount compensated, insured area in hectares and the number of producers benefited per crop.
However, they were not the only crops affected, with a minimal difference in insured area and large amounts of compensation are the crops of potato, quinoa, and broad bean, all this at the regional level of the data recorded in the 2010-2011 agricultural campaign. 

This year, according to MINAGRI, there are 9 regions that show a decrease in their planting intentions in these same crops: Amazonas, Arequipa, Huanuco, Ica, Junin, Metropolitan Lima, Loreto, Pasco, and Puno, which together would plant 666,130 hectares (32 percent of the total planting intentions for the new 2022/2023 crop year), decreasing by 4.3 percent (30,160 hectares less) compared to the average of the last five crop years\cite{DEAI}.

\begin{table}[ht]
    \centering
    \begin{tabular}{|c|c|c|}
        \toprule
        PROVINCE & {SUPER ASEGUR} & {AMOUNT INDEM S/.} \\
        \midrule
        CHUCUITO & 1817 & 726800 \\
        EL COLLAO & 3910 & 1564000 \\
        AZANGARO & 3019 & 1207600 \\
        CARABAYA & 700 & 280000 \\
        HUANCANE & 1399 & 559600 \\
        LAMPA & 932 & 372800 \\
        MELGAR & 1455 & 582000 \\
        SAN ROMAN & 766 & 306400 \\
        YUNGUYO & 840 & 336000 \\
        \midrule
        TOTAL & 14838 & 5935200 \\
        \bottomrule
    \end{tabular}
    \caption{REGION PUNO: CATASTROPHIC AGRARIAN INSURANCE BY PROVINCE, YEAR 2011}
\end{table}

In this case, the consolidated figures are at the provincial level, with the total insured area in hectares together with the amount indemnified per province.
The most representative provinces are El Collao (S/.1564000.00) and Azangaro (S/.1207600.00), which means that these provinces had more crop losses, followed by the other provinces.
\subsection{Graphics and visualizations:}

The bar chart shows the distribution of insured area by crop for each crop year in the Puno region. It can be seen that the potato crop has the largest insured area in general.
The summary table by province shows that the Aymaraes province has the largest insured area, the highest indemnity amount, and the largest number of beneficiary producers.
The scatter plot shows the relationship between the insured area and the indemnity amount per crop. A positive trend is evident, where crops with greater insured areas also receive higher compensation.

\begin{figure}[h!]
    \centering
    \includegraphics[width=8cm,height=6cm]{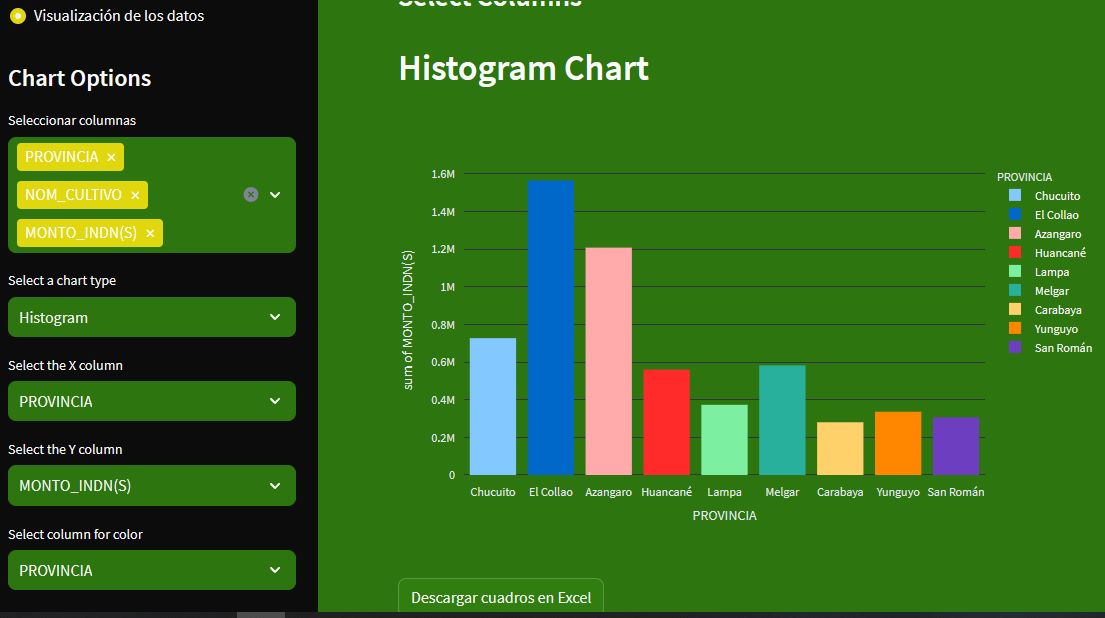}
    \caption{ SAC application in Streamlit, section graphics}
\end{figure}

The analysis of the results shows that agricultural insurance in the Puno region plays an important role in protecting farmers against losses caused by adverse weather events. The results highlight the need to focus resources and compensation on the most affected crops and in the provinces with the highest agricultural activity.\cite{malpartida2018desarrollo}

It also emphasizes the need to promote agricultural practices that are adaptable to climate change and to strengthen risk prevention and mitigation strategies to reduce losses and improve food security in the region.
\subsection{Time Series:}
\subsubsection{Estimation of the Identified Model}

An ARIMA (0,1,1) (0,1,1) model has been identified, indicating a non-stationary model with 0 in autoregressive order, 1 in difference order, and 1 in moving average order, and a seasonal model with 0 in autoregressive order, 1 in difference order and 1 in moving average order.

The final parameter estimates are as follows:

\begin{center}
\begin{tabular}{lccccc}
\hline
Tipo & Coef & SE Coef & Valor T & Valor p \\
\hline
MA   1 & 0.8922 & 0.0235 & 37.92 & 0.000 \\
SMA  12 & 0.9235 & 0.0282 & 32.70 & 0.000 \\
\hline
\end{tabular}
\end{center}

Modified Box-Pierce (Ljung-Box) chi-square statistics have been calculated for different lags:

\begin{center}
\begin{tabular}{ccccc}
\hline
Desfase & 12 & 24 & 36 & 48 \\

Chi-2 & 11.63 & 26.15 &36.32 &47.47 \\

GL & 10 &22 &34 &46 \\

Valor p & 0.311 &0.245 &0.361 &0.413 \\
\hline
\end{tabular}
\end{center}

The parameter $w$ is estimated by the product:
\[ w_1 \times w_{12} = 0.8922 \times 0.9235 = 0.8239467 \]

The estimated model is:
\[ \hat{Y}_t = Y_{t-1}+Y_{t-12}-Y_{t-13} - \]
\[ \hat0.8922e_{t-1}- 0.9235e_{t-12} + 0.8239e_{t-13} \]

\subsubsection{Model Validation}
1) Two conditions are met within the parameter estimates: for MA (0.8922) and SMA (0.9235) are less than 1 and the $p$ value is less than 0.05 in both cases, indicating that the MA and SMA results are significant.

2)The next condition that is met is that the $p$ values of the modified Box-Pierce (Ljung-Box) chi-square statistic are greater than the significance level of 0.05 and if such condition is met (0.311), (0.245), (0.361) and (0.413).

3)Likewise, the chi-square is present.
\begin{figure}[h!]
    \centering
    \includegraphics[width=8cm,height=5cm]{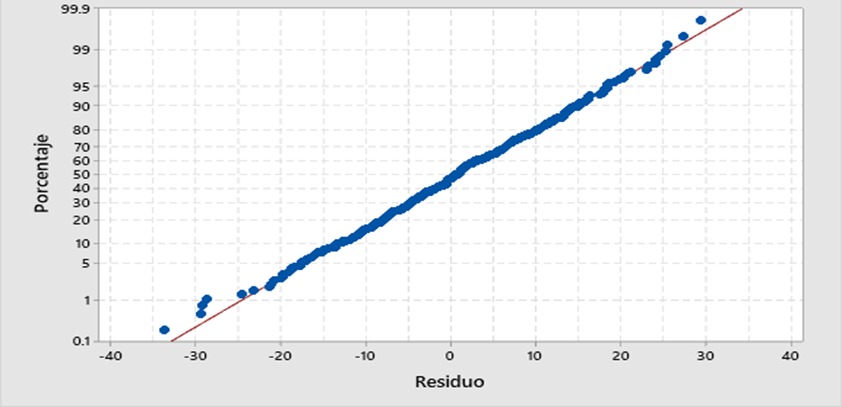}
    \caption{ Normal probability plot}
\end{figure}
This figure shows that the data after applying the model follow a normal distribution, with the exception of some data, but in general, it is assumed that the proposed model is adequate.
 \section{ DISCUSSION AND CONCLUSION}
From the tests performed, it was possible to distinguish a clear relationship between a person's technical knowledge and the need for documentation for equipment configuration\cite{torres2022desarrollo}. In this study, a user-friendly interface based on the principles of usability, aesthetics, and user satisfaction was demonstrated. The results obtained show that the implemented interface was well received by the participants, who reported high levels of satisfaction and an overall positive experience.

The user-friendly interface will work in terms of facilitating user interaction with the system, requiring a smooth and seamless experience\cite{jorquera2019evaluacion}. The intuitive layout of interface elements, attractive visual design, and easy navigation contributed to a positive user experience. This evidence supports previous studies highlighting the importance of usability and aesthetics in user perception and satisfaction with an interface\cite{hernandez2023estrategia}.

In addition, it was shown that participants were able to complete the ordered tasks easily and quickly. This indicates that the user-friendly interface was effective in supporting user efficiency in interacting with the system\cite{encalada2018uso}. 
In conclusion, the results of this study support the importance of developing user-friendly interfaces that meet the principles of usability, aesthetics, and user satisfaction. The implementation of a user-friendly interface can improve the user experience, facilitate interaction with the system, and increase overall satisfaction\cite{rodriguez2017arquitectura}. However, it is essential to continue to research and refine adaptive interface design techniques to adapt to changing user needs and expectations.

In conclusion, the data analysis demonstrates the effectiveness of agricultural insurance in the Puno region. The results obtained provided valuable information for decision-making in the design of agricultural policies, the use of resources, and the implementation of risk mitigation strategies. It also highlights the need to continue monitoring and evaluating the agricultural insurance program to improve its scope and benefits for farmers.
\newpage
\bibliography{referencias}

\begin{thebibliography}{}

\bibitem [\protect \citeauthoryear {%
An{\v{z}}el%
, Heider%
\BCBL {}\ \BBA {} Hattab%
}{%
An{\v{z}}el%
\ \protect \BOthers {.}}{%
{\protect \APACyear {2022}}%
}]{%
anvzel2022movis}
\APACinsertmetastar {%
anvzel2022movis}%
\begin{APACrefauthors}%
An{\v{z}}el, A.%
, Heider, D.%
\BCBL {}\ \BBA {} Hattab, G.%
\end{APACrefauthors}%
\unskip\
\newblock
\APACrefYearMonthDay{2022}{}{}.
\newblock
{\BBOQ}\APACrefatitle {MOVIS: A multi-omics software solution for multi-modal
  time-series clustering, embedding, and visualizing tasks} {Movis: A
  multi-omics software solution for multi-modal time-series clustering,
  embedding, and visualizing tasks}.{\BBCQ}
\newblock
\APACjournalVolNumPages{Computational and Structural Biotechnology
  Journal}{20}{}{1044--1055}.
\PrintBackRefs{\CurrentBib}

\bibitem [\protect \citeauthoryear {%
Bula%
}{%
Bula%
}{%
{\protect \APACyear {2020}}%
}]{%
bula2020importance}
\APACinsertmetastar {%
bula2020importance}%
\begin{APACrefauthors}%
Bula, A\BPBI O.%
\end{APACrefauthors}%
\unskip\
\newblock
\APACrefYearMonthDay{2020}{}{}.
\newblock
{\BBOQ}\APACrefatitle {Importance of agriculture in socio-economic development}
  {Importance of agriculture in socio-economic development}.{\BBCQ}
\newblock

\PrintBackRefs{\CurrentBib}

\bibitem [\protect \citeauthoryear {%
Carlos David~Felipe%
\ \BBA {} Gabriel~Andr{\'e}s%
}{%
Carlos David~Felipe%
\ \BBA {} Gabriel~Andr{\'e}s%
}{%
{\protect \APACyear {2022}}%
}]{%
carlos2022aplicativo}
\APACinsertmetastar {%
carlos2022aplicativo}%
\begin{APACrefauthors}%
Carlos David~Felipe, R\BPBI C.%
\BCBT {}\ \BBA {} Gabriel~Andr{\'e}s, M\BPBI S.%
\end{APACrefauthors}%
\unskip\
\newblock
\APACrefYearMonthDay{2022}{}{}.
\newblock
{\BBOQ}\APACrefatitle {Web application for the quick and safe selection of a
  suitable cargo vehicle for a trip} {Web application for the quick and safe
  selection of a suitable cargo vehicle for a trip}.{\BBCQ}
\newblock

\PrintBackRefs{\CurrentBib}

\bibitem [\protect \citeauthoryear {%
Castillo~Chiang%
, Andrade~Chang%
\BCBL {}\ \protect \BOthers {.}}{%
Castillo~Chiang%
\ \protect \BOthers {.}}{%
{\protect \APACyear {2018}}%
}]{%
castillo2018diseno}
\APACinsertmetastar {%
castillo2018diseno}%
\begin{APACrefauthors}%
Castillo~Chiang, J\BPBI A.%
, Andrade~Chang, C\BPBI X.%
\BCBL {}\ \BOthersPeriod {.}\end{APACrefauthors}%
\unskip\
\newblock
\APACrefYear{2018}.
\unskip\
\newblock
\APACrefbtitle {Design and implementation of a device for presentation of
  multimedia material using low-cost hardware and free software} {Design and
  implementation of a device for presentation of multimedia material using
  low-cost hardware and free software}\ \APACtypeAddressSchool {{B.S.}
  thesis}{}{}.
\PrintBackRefs{\CurrentBib}

\bibitem [\protect \citeauthoryear {%
Chaure~Cordero%
}{%
Chaure~Cordero%
}{%
{\protect \APACyear {2022}}%
}]{%
chaure2022forecaster}
\APACinsertmetastar {%
chaure2022forecaster}%
\begin{APACrefauthors}%
Chaure~Cordero, P.%
\end{APACrefauthors}%
\unskip\
\newblock
\APACrefYearMonthDay{2022}{}{}.
\newblock
{\BBOQ}\APACrefatitle {FORECASTER. APLICACI{\'O}N WEB SHINY PARA EL
  AN{\'A}LISIS Y EXPERIMENTACI{\'O}N DE MODELOS DE SERIES TEMPORALES}
  {Forecaster. aplicaci{\'o}n web shiny para el an{\'a}lisis y
  experimentaci{\'o}n de modelos de series temporales}.{\BBCQ}
\newblock

\PrintBackRefs{\CurrentBib}

\bibitem [\protect \citeauthoryear {%
Chura%
\ \BBA {} Mamani%
}{%
Chura%
\ \BBA {} Mamani%
}{%
{\protect \APACyear {2019}}%
}]{%
chura2019necesidades}
\APACinsertmetastar {%
chura2019necesidades}%
\begin{APACrefauthors}%
Chura, F\BPBI C.%
\BCBT {}\ \BBA {} Mamani, J\BPBI I.%
\end{APACrefauthors}%
\unskip\
\newblock
\APACrefYearMonthDay{2019}{}{}.
\newblock
{\BBOQ}\APACrefatitle {Needs of agricultural and livestock technology in the
  Accaso-Puno Peruvian community} {Needs of agricultural and livestock
  technology in the accaso-puno peruvian community}.{\BBCQ}
\newblock
\APACjournalVolNumPages{Ciencia \& Desarrollo}{}{24}{27--37}.
\PrintBackRefs{\CurrentBib}

\bibitem [\protect \citeauthoryear {%
Coppiano~Mar{\'\i}n%
\ \BBA {} Herrera~Vargas%
}{%
Coppiano~Mar{\'\i}n%
\ \BBA {} Herrera~Vargas%
}{%
{\protect \APACyear {2022}}%
}]{%
coppiano2022desarrollo}
\APACinsertmetastar {%
coppiano2022desarrollo}%
\begin{APACrefauthors}%
Coppiano~Mar{\'\i}n, A\BPBI D.%
\BCBT {}\ \BBA {} Herrera~Vargas, C\BPBI J.%
\end{APACrefauthors}%
\unskip\
\newblock
\APACrefYear{2022}.
\unskip\
\newblock
\APACrefbtitle {Development of a web application based on support vector
  machines (svm) of supervised learning for the prediction of crop
  recommendation using environmental data for agroecological farms in the
  canton of La Mancha, province of Cotopaxi.} {Development of a web application
  based on support vector machines (svm) of supervised learning for the
  prediction of crop recommendation using environmental data for agroecological
  farms in the canton of la mancha, province of cotopaxi.}\
  \APACtypeAddressSchool {{B.S.} thesis}{}{}.
\unskip\
\newblock
\APACaddressSchool {}{Ecuador: La Mana: Universidad T{\'e}cnica de Cotopaxi
  (UTC)}.
\PrintBackRefs{\CurrentBib}

\bibitem [\protect \citeauthoryear {%
\APACcitebtitle {DEAI: The fastest way to build custom ML tools}}{%
\APACcitebtitle {DEAI: The fastest way to build custom ML tools}}{%
{\protect \APACyear {2023}}%
}]{%
DEAI}
\APACinsertmetastar {%
DEAI}%
\APACrefbtitle {DEAI: The fastest way to build custom ML tools.} {Deai: The
  fastest way to build custom ml tools.}
\newblock
\APACrefYearMonthDay{2023}{}{}.
\newblock
\APAChowpublished {\url{https://www.agropuno.gob.pe/informacion-estadistica/}}.
\newblock
\APACrefnote{Accedido el 2 de julio de 2023}
\PrintBackRefs{\CurrentBib}

\bibitem [\protect \citeauthoryear {%
Dominguez%
, Eijo%
\BCBL {}\ \BBA {} Felix%
}{%
Dominguez%
\ \protect \BOthers {.}}{%
{\protect \APACyear {2022}}%
}]{%
dominguez2022analisis}
\APACinsertmetastar {%
dominguez2022analisis}%
\begin{APACrefauthors}%
Dominguez, L.%
, Eijo, G.%
\BCBL {}\ \BBA {} Felix, S.%
\end{APACrefauthors}%
\unskip\
\newblock
\APACrefYearMonthDay{2022}{}{}.
\newblock
{\BBOQ}\APACrefatitle {Analysis of news about citizen security in social
  networks} {Analysis of news about citizen security in social
  networks}.{\BBCQ}
\newblock

\PrintBackRefs{\CurrentBib}

\bibitem [\protect \citeauthoryear {%
Encalada~D{\'\i}az%
\ \BBA {} Delgado~Alva%
}{%
Encalada~D{\'\i}az%
\ \BBA {} Delgado~Alva%
}{%
{\protect \APACyear {2018}}%
}]{%
encalada2018uso}
\APACinsertmetastar {%
encalada2018uso}%
\begin{APACrefauthors}%
Encalada~D{\'\i}az, I\BPBI A.%
\BCBT {}\ \BBA {} Delgado~Alva, R.%
\end{APACrefauthors}%
\unskip\
\newblock
\APACrefYearMonthDay{2018}{}{}.
\newblock
{\BBOQ}\APACrefatitle {The use of the educational software cuadernia in the
  teaching-learning process and in the academic performance of mathematics of
  the students of the 5th year of secondary school of the educational
  institution N{textordmasculine} 5143 school of talents Callao 2015} {The use
  of the educational software cuadernia in the teaching-learning process and in
  the academic performance of mathematics of the students of the 5th year of
  secondary school of the educational institution n{textordmasculine} 5143
  school of talents callao 2015}.{\BBCQ}
\newblock

\PrintBackRefs{\CurrentBib}

\bibitem [\protect \citeauthoryear {%
Falconi~Sarmiento%
}{%
Falconi~Sarmiento%
}{%
{\protect \APACyear {2017}}%
}]{%
falconi2017situacion}
\APACinsertmetastar {%
falconi2017situacion}%
\begin{APACrefauthors}%
Falconi~Sarmiento, G\BPBI A.%
\end{APACrefauthors}%
\unskip\
\newblock
\APACrefYearMonthDay{2017}{}{}.
\newblock
{\BBOQ}\APACrefatitle {Situation of agrarian insurance in Peru: catastrophic
  agricultural insurance} {Situation of agrarian insurance in peru:
  catastrophic agricultural insurance}.{\BBCQ}
\newblock

\PrintBackRefs{\CurrentBib}

\bibitem [\protect \citeauthoryear {%
Hern{\'a}ndez~Barajas%
}{%
Hern{\'a}ndez~Barajas%
}{%
{\protect \APACyear {2023}}%
}]{%
hernandez2023estrategia}
\APACinsertmetastar {%
hernandez2023estrategia}%
\begin{APACrefauthors}%
Hern{\'a}ndez~Barajas, D\BPBI C.%
\end{APACrefauthors}%
\unskip\
\newblock
\APACrefYear{2023}.
\unskip\
\newblock
\APACrefbtitle {Monetization strategy in UX user experience design for startups
  on web platforms and mobile applications} {Monetization strategy in ux user
  experience design for startups on web platforms and mobile applications}\
  \APACtypeAddressSchool {\BUPhD}{}{}.
\unskip\
\newblock
\APACaddressSchool {}{Universidad Aut{\'o}noma de Bucaramanga UNAB}.
\PrintBackRefs{\CurrentBib}

\bibitem [\protect \citeauthoryear {%
Jorquera~Soto%
\ \protect \BOthers {.}}{%
Jorquera~Soto%
\ \protect \BOthers {.}}{%
{\protect \APACyear {2019}}%
}]{%
jorquera2019evaluacion}
\APACinsertmetastar {%
jorquera2019evaluacion}%
\begin{APACrefauthors}%
Jorquera~Soto, V\BPBI E.%
\BCBT {}\ \BOthersPeriod {.}
\end{APACrefauthors}%
\unskip\
\newblock
\APACrefYearMonthDay{2019}{}{}.
\newblock
{\BBOQ}\APACrefatitle {Evaluation and proposal for improvement of the online
  catalog interface (OPAC) of the libraries of the Universidad Adolfo Ibáñez,
  Chile.} {Evaluation and proposal for improvement of the online catalog
  interface (opac) of the libraries of the universidad adolfo ibáñez,
  chile.}{\BBCQ}
\newblock

\PrintBackRefs{\CurrentBib}

\bibitem [\protect \citeauthoryear {%
Malpartida~Calder{\'o}n%
}{%
Malpartida~Calder{\'o}n%
}{%
{\protect \APACyear {2018}}%
}]{%
malpartida2018desarrollo}
\APACinsertmetastar {%
malpartida2018desarrollo}%
\begin{APACrefauthors}%
Malpartida~Calder{\'o}n, G\BPBI J.%
\end{APACrefauthors}%
\unskip\
\newblock
\APACrefYearMonthDay{2018}{}{}.
\newblock
{\BBOQ}\APACrefatitle {Development of agricultural insurance in the Peruvian
  market} {Development of agricultural insurance in the peruvian
  market}.{\BBCQ}
\newblock

\PrintBackRefs{\CurrentBib}

\bibitem [\protect \citeauthoryear {%
McKinney%
}{%
McKinney%
}{%
{\protect \APACyear {2010}}%
}]{%
McKinney:2010}
\APACinsertmetastar {%
McKinney:2010}%
\begin{APACrefauthors}%
McKinney, W.%
\end{APACrefauthors}%
\unskip\
\newblock
\APACrefYearMonthDay{2010}{}{}.
\newblock
{\BBOQ}\APACrefatitle {Data Structures for Statistical Computing in Python}
  {Data structures for statistical computing in python}.{\BBCQ}
\newblock
\BIn{} \APACrefbtitle {Proceedings of the 9th Python in Science Conference}
  {Proceedings of the 9th python in science conference}\ (\BPG~51-56).
\newblock
\APACrefnote{Recuperado de
  \url{https://conference.scipy.org/proceedings/scipy2010/mckinney.html}}
\PrintBackRefs{\CurrentBib}

\bibitem [\protect \citeauthoryear {%
Olascoaga~Mouchard%
}{%
Olascoaga~Mouchard%
}{%
{\protect \APACyear {2023}}%
}]{%
olascoaga2023gestion}
\APACinsertmetastar {%
olascoaga2023gestion}%
\begin{APACrefauthors}%
Olascoaga~Mouchard, J\BPBI A.%
\end{APACrefauthors}%
\unskip\
\newblock
\APACrefYearMonthDay{2023}{}{}.
\newblock
{\BBOQ}\APACrefatitle {Financial risk management to face the effects of climate
  change in Peruvian agriculture. Case of Mala Valley Lima Region} {Financial
  risk management to face the effects of climate change in peruvian
  agriculture. case of mala valley lima region}.{\BBCQ}
\newblock

\PrintBackRefs{\CurrentBib}

\bibitem [\protect \citeauthoryear {%
Pineda~Torres%
\ \BBA {} Iba{\~n}ez~Lozano%
}{%
Pineda~Torres%
\ \BBA {} Iba{\~n}ez~Lozano%
}{%
{\protect \APACyear {2023}}%
}]{%
pineda2023diseno}
\APACinsertmetastar {%
pineda2023diseno}%
\begin{APACrefauthors}%
Pineda~Torres, A\BPBI F.%
\BCBT {}\ \BBA {} Iba{\~n}ez~Lozano, M\BPBI S.%
\end{APACrefauthors}%
\unskip\
\newblock
\APACrefYearMonthDay{2023}{}{}.
\newblock
{\BBOQ}\APACrefatitle {Design and implementation of a device to collect
  demographic information such as age and gender through the use of computer
  vision} {Design and implementation of a device to collect demographic
  information such as age and gender through the use of computer
  vision}.{\BBCQ}
\newblock

\PrintBackRefs{\CurrentBib}

\bibitem [\protect \citeauthoryear {%
Rodr{\'\i}guez~Castilla%
, Gonz{\'a}lez~Hern{\'a}ndez%
\BCBL {}\ \BBA {} P{\'e}rez~Gonz{\'a}lez%
}{%
Rodr{\'\i}guez~Castilla%
\ \protect \BOthers {.}}{%
{\protect \APACyear {2017}}%
}]{%
rodriguez2017arquitectura}
\APACinsertmetastar {%
rodriguez2017arquitectura}%
\begin{APACrefauthors}%
Rodr{\'\i}guez~Castilla, L.%
, Gonz{\'a}lez~Hern{\'a}ndez, D\BPBI L.%
\BCBL {}\ \BBA {} P{\'e}rez~Gonz{\'a}lez, Y.%
\end{APACrefauthors}%
\unskip\
\newblock
\APACrefYearMonthDay{2017}{}{}.
\newblock
{\BBOQ}\APACrefatitle {From information architecture to user experience: Their
  interrelation in software development at the University of Computer Science}
  {From information architecture to user experience: Their interrelation in
  software development at the university of computer science}.{\BBCQ}
\newblock
\APACjournalVolNumPages{E-Ciencias de la Informaci{\'o}n}{7}{1}{1--24}.
\PrintBackRefs{\CurrentBib}

\bibitem [\protect \citeauthoryear {%
Rodr{\'\i}guez~Rodr{\'\i}guez%
}{%
Rodr{\'\i}guez~Rodr{\'\i}guez%
}{%
{\protect \APACyear {2017}}%
}]{%
rodriguez2017factores}
\APACinsertmetastar {%
rodriguez2017factores}%
\begin{APACrefauthors}%
Rodr{\'\i}guez~Rodr{\'\i}guez, M.%
\end{APACrefauthors}%
\unskip\
\newblock
\APACrefYearMonthDay{2017}{}{}.
\newblock
{\BBOQ}\APACrefatitle {Factors that Determine the Jurisdictional Preference for
  the Resolution of Conflicts of Interest of the Farmers of Carabaya--Puno
  2014} {Factors that determine the jurisdictional preference for the
  resolution of conflicts of interest of the farmers of carabaya--puno
  2014}.{\BBCQ}
\newblock

\PrintBackRefs{\CurrentBib}

\bibitem [\protect \citeauthoryear {%
Romero%
}{%
Romero%
}{%
{\protect \APACyear {2017}}%
}]{%
romero2017seguro}
\APACinsertmetastar {%
romero2017seguro}%
\begin{APACrefauthors}%
Romero, R\BPBI B.%
\end{APACrefauthors}%
\unskip\
\newblock
\APACrefYearMonthDay{2017}{}{}.
\newblock
{\BBOQ}\APACrefatitle {Agrarian Insurance in the province of Corrientes}
  {Agrarian insurance in the province of corrientes}.{\BBCQ}
\newblock
\BIn{} \APACrefbtitle {IV Congreso Nacional de Derecho Agrario Provincial
  (Salta, 2017).} {Iv congreso nacional de derecho agrario provincial (salta,
  2017).}
\PrintBackRefs{\CurrentBib}

\bibitem [\protect \citeauthoryear {%
\APACcitebtitle {Streamlit: The fastest way to build custom ML tools}}{%
\APACcitebtitle {Streamlit: The fastest way to build custom ML tools}}{%
{\protect \APACyear {2023}}%
}]{%
Streamlit}
\APACinsertmetastar {%
Streamlit}%
\APACrefbtitle {Streamlit: The fastest way to build custom ML tools.}
  {Streamlit: The fastest way to build custom ml tools.}
\newblock
\APACrefYearMonthDay{2023}{}{}.
\newblock
\APAChowpublished {\url{https://streamlit.io/}}.
\newblock
\APACrefnote{Accedido el 2 de julio de 2023}
\PrintBackRefs{\CurrentBib}

\bibitem [\protect \citeauthoryear {%
Torres~Castillo%
}{%
Torres~Castillo%
}{%
{\protect \APACyear {2022}}%
}]{%
torres2022desarrollo}
\APACinsertmetastar {%
torres2022desarrollo}%
\begin{APACrefauthors}%
Torres~Castillo, E\BPBI D.%
\end{APACrefauthors}%
\unskip\
\newblock
\APACrefYear{2022}.
\unskip\
\newblock
\APACrefbtitle {Development of a mobile application for teaching guitar and
  Ukulele with augmented reality. Case study: Import Music} {Development of a
  mobile application for teaching guitar and ukulele with augmented reality.
  case study: Import music}\ \APACtypeAddressSchool {{B.S.} thesis}{}{}.
\unskip\
\newblock
\APACaddressSchool {}{PUCE-Quito}.
\PrintBackRefs{\CurrentBib}

\bibitem [\protect \citeauthoryear {%
Vinelli%
}{%
Vinelli%
}{%
{\protect \APACyear {2015}}%
}]{%
vinelli2015informe}
\APACinsertmetastar {%
vinelli2015informe}%
\begin{APACrefauthors}%
Vinelli, M.%
\end{APACrefauthors}%
\unskip\
\newblock
\APACrefYearMonthDay{2015}{}{}.
\newblock
\APACrefbtitle {SAC Final Report.} {Sac final report.}
\newblock
\APAChowpublished {Lima: Dirección General de Negocios Agrarios - MINAGRI}.
\PrintBackRefs{\CurrentBib}

\bibitem [\protect \citeauthoryear {%
Wakchaure%
, Patle%
\BCBL {}\ \BBA {} Mahindrakar%
}{%
Wakchaure%
\ \protect \BOthers {.}}{%
{\protect \APACyear {2023}}%
}]{%
wakchaure2023application}
\APACinsertmetastar {%
wakchaure2023application}%
\begin{APACrefauthors}%
Wakchaure, M.%
, Patle, B.%
\BCBL {}\ \BBA {} Mahindrakar, A.%
\end{APACrefauthors}%
\unskip\
\newblock
\APACrefYearMonthDay{2023}{}{}.
\newblock
{\BBOQ}\APACrefatitle {Application of AI techniques and robotics in
  agriculture: A review} {Application of ai techniques and robotics in
  agriculture: A review}.{\BBCQ}
\newblock
\APACjournalVolNumPages{Artificial Intelligence in the Life
  Sciences}{}{}{100057}.
\PrintBackRefs{\CurrentBib}

\end{thebibliography}
\end{document}